\documentclass[runningheads]{llncs}
\usepackage[T1]{fontenc}
\usepackage{graphicx}
\usepackage{amsmath}
\usepackage{tikz}
\usepackage{pgfplots}
\usepackage{multirow}
\pgfplotsset{compat=1.18} 
\begin{document}

\title{Towards Building Secure UAV Navigation with FHE-aware Knowledge Distillation}
%
%
\author{Arjun Ramesh Kaushik\inst{1} \and
Charanjit Jutla\inst{2} \and Nalini Ratha\inst{1}}
%
%

\institute{University at Buffalo, The State University of New York, USA \and IBM Research, USA\\
\email{\{kaushik3,nratha\}@buffalo.edu, csjutla@us.ibm.com}}

\maketitle              
\begin{abstract}
In safeguarding mission-critical systems, such as Unmanned Aerial Vehicles (UAVs), preserving the privacy of path trajectories during navigation is paramount. While the combination of Reinforcement Learning (RL) and Fully Homomorphic Encryption (FHE) holds promise, the computational overhead of FHE presents a significant challenge. This paper proposes an innovative approach that leverages Knowledge Distillation to enhance the practicality of secure UAV navigation. By integrating RL and FHE, our framework addresses vulnerabilities to adversarial attacks while enabling real-time processing of encrypted UAV camera feeds, ensuring data security. To mitigate FHE's latency, Knowledge Distillation is employed to compress the network, resulting in an impressive 18x speedup without compromising performance, as evidenced by an R-squared score of 0.9499 compared to the original model's score of 0.9631. Our methodology underscores the feasibility of processing encrypted data for UAV navigation tasks, emphasizing security alongside performance efficiency and timely processing. These findings pave the way for deploying autonomous UAVs in sensitive environments, bolstering their resilience against potential security threats.

\end{abstract}
\section{Introduction}
In recent years, the integration of autonomous Unmanned Aerial Vehicles (UAVs) has revolutionized various industries, offering unparalleled capabilities in surveillance, reconnaissance, disaster response, and product delivery \cite{drones6060147}. However, ensuring secure navigation of UAVs, particularly in critical scenarios, has become a paramount concern due to the inherent vulnerabilities associated with Deep Learning (DL) techniques and potential adversarial attacks \cite{makdad_survey_drone}\cite{guo_wang_uav_secure}. While previous research has made strides in enhancing UAV security  \cite{drone_nav_cai}\cite{she_drone}, the computational demands of existing solutions often render them impractical for real-world deployment. This paper addresses the pressing need for a secure and feasible architecture for UAV navigation.

\begin{figure*}[!ht]
\centerline{\includegraphics[width=1\textwidth]{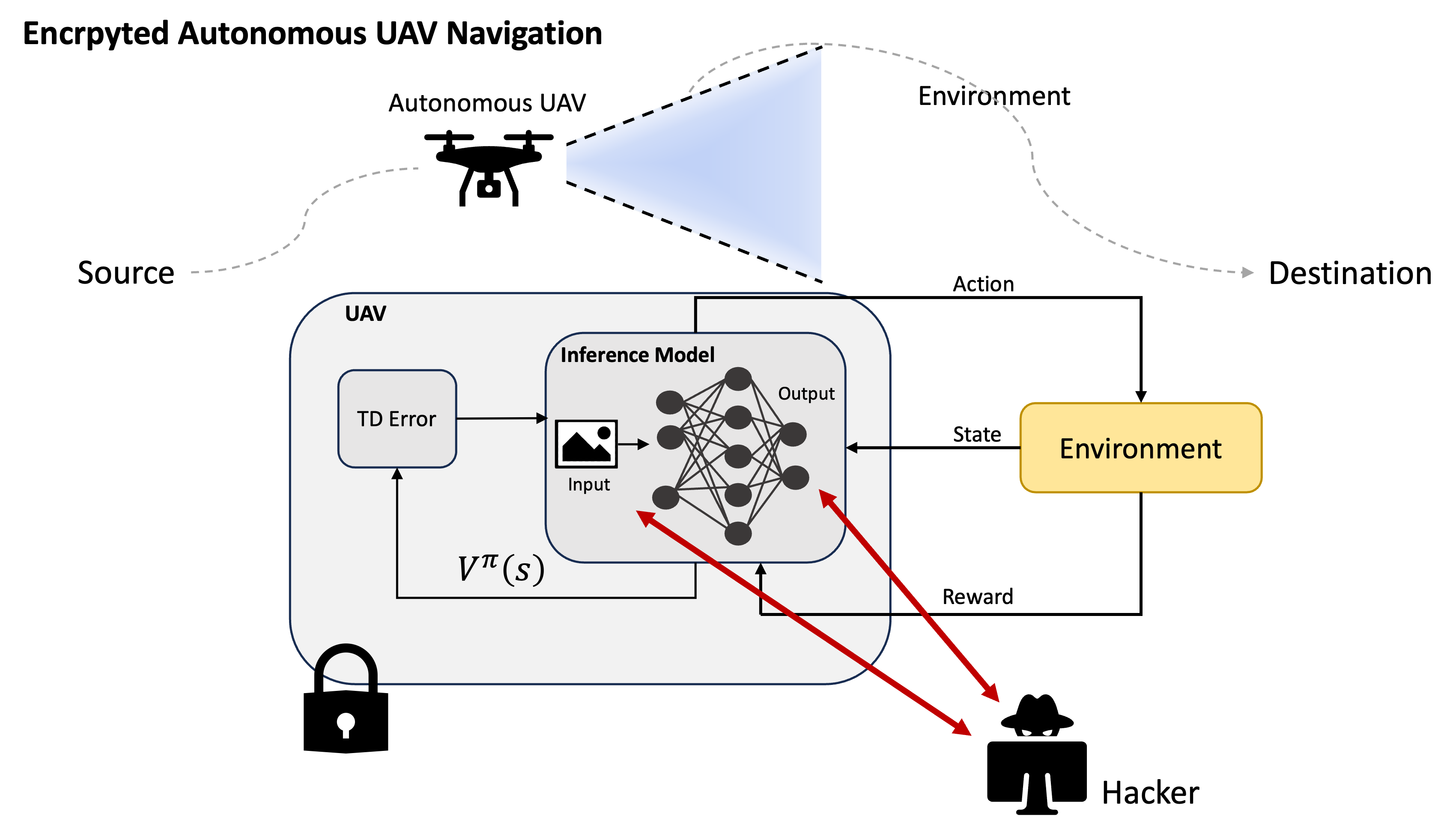}}
\caption{\textbf{Overview:} In an ordinary scenario the UAV is vulnerable to attacks, as the attacker can directly steal the information. FHE-encrypted input and inference prevent this. But, currently, FHE is computationally infeasible.}
\label{intro_fig}
\end{figure*}

\begin{figure*}[!ht]
\centerline{\includegraphics[width=1\textwidth]{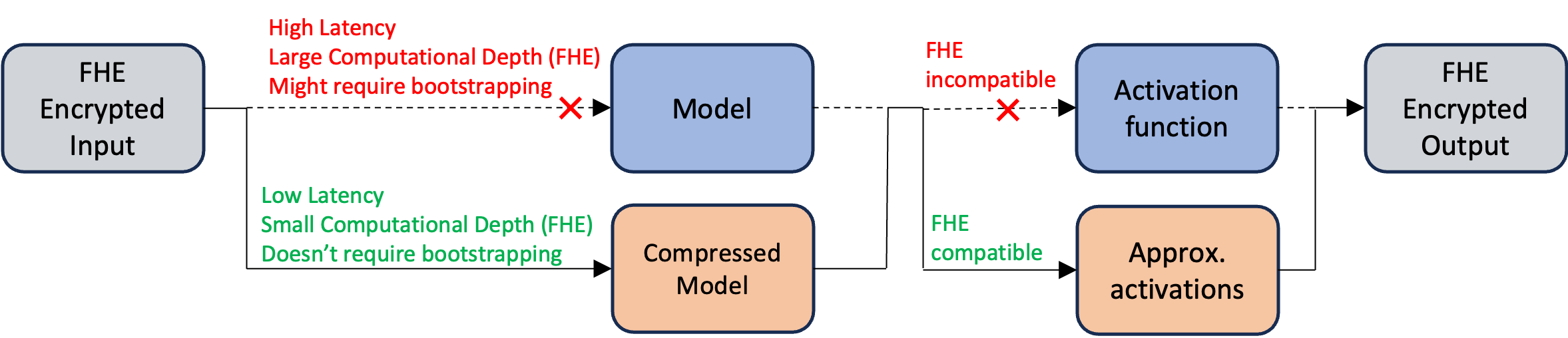}}
\caption{An overview of the need for an FHE optimized model.}
\label{fhe}
\end{figure*}

\begin{figure*}[!ht]
\centerline{\includegraphics[width=1\textwidth]{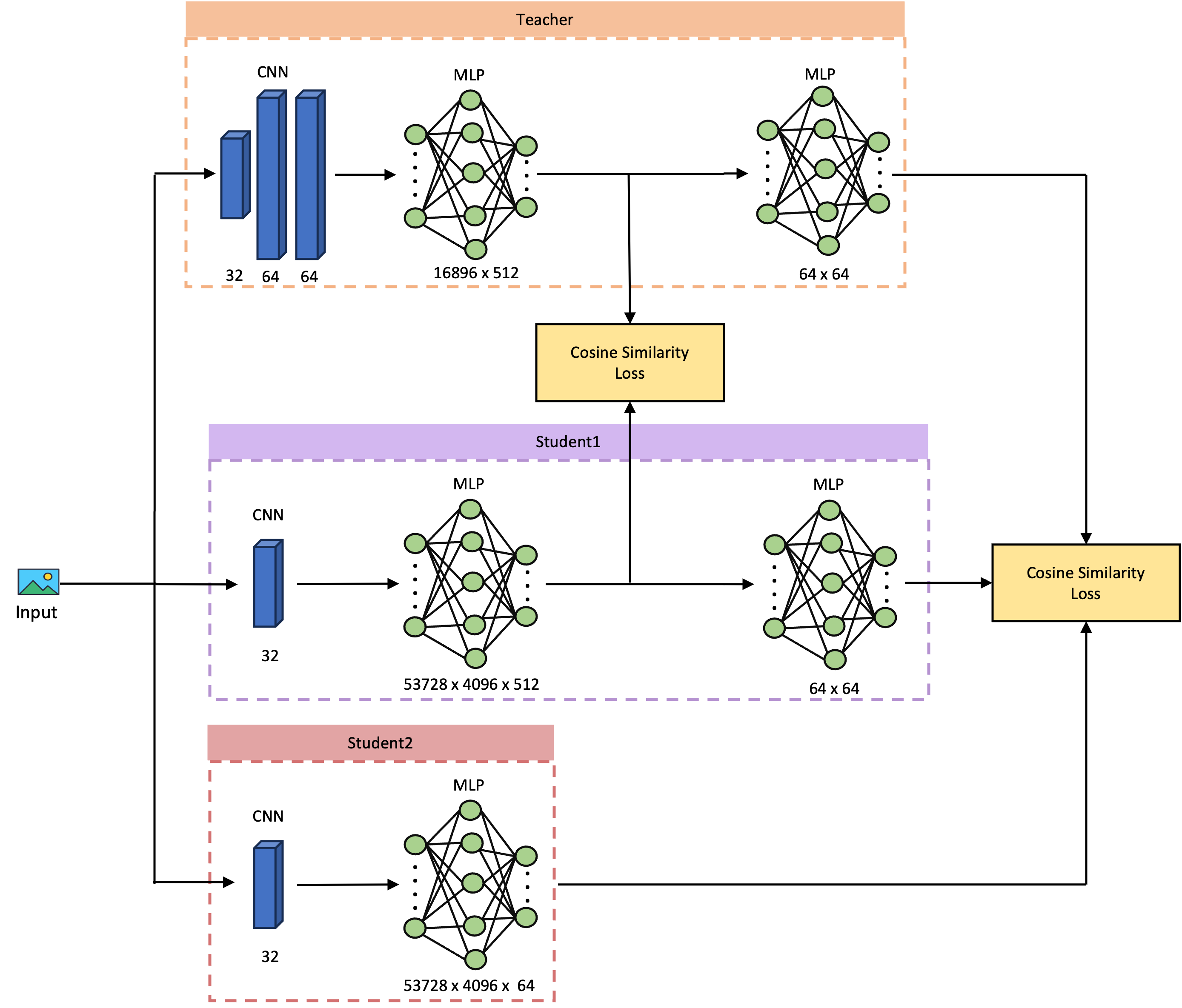}}
\caption{We propose a smaller model through Knowledge Distillation to suit FHE needs while maintaining security and accuracy.}
\label{kd}
\end{figure*}

While traditional approaches to UAV navigation have relied on vision-based systems incorporating visual mapping, obstacle detection, and path planning \cite{doi:10.1080/10095020.2017.1420509}, recent advancements have shifted towards leveraging Deep Learning and Reinforcement Learning methodologies \cite{8993742,8600371,9729807}. In response to the increasing importance of security, recent works have explored various security schemes \cite{drone_nav_cai,LinHAE,secure_survey}. However, many existing solutions either prioritize maximum security at the expense of computational feasibility or offer compromised security with practical implementation. Our contribution introduces a secure Reinforcement Learning framework, utilizing the Actor-Critic policy within the Proximal Policy Optimization (PPO) algorithm, capable of seamlessly operating on encrypted real-time video feeds captured by UAV cameras, while remaining resilient to adversarial attacks (Fig. \ref{intro_fig}). Building upon prior research \cite{drone_nav_cai}, we present a significantly more feasible architecture in terms of computational efficiency.

In the subsequent sections, we provide a comprehensive overview of how each component of our deep learning model is uniquely adapted to handle encrypted data. Key aspects of our approach include transforming convolutional layers into spectral domain operations, employing generalized matrix multiplication in fully connected layers, and customizing activation functions for the FHE domain through polynomial approximations and comparators. Additionally, navigational steps are extracted through a neural network trained to replicate the OpenAI Gym library. Despite the maximum security provided by FHE, its computational overhead remains significant even after adaptation. To address this challenge, we propose a smaller model through Knowledge Distillation, ensuring feasibility within the FHE framework. Importantly, our research demonstrates the minimal loss of accuracy when mapping teacher and student models to the FHE domain, validating the feasibility of processing encrypted data for UAV navigation tasks.

This work not only addresses immediate security concerns associated with UAVs, but also lays the groundwork for a new era in autonomous aerial systems. By prioritizing security and privacy through FHE integration, our approach opens avenues for deploying UAVs in sensitive domains where data confidentiality is paramount. The implications extend to applications in military operations, surveillance, and disaster response, where enhanced security measures are essential for the successful execution of critical missions.

\section{Threat Model}

Unmanned Aerial Vehicles (UAVs) deployed in critical scenarios are exposed to various adversarial threats, including (i) Data Poisoning \cite{uav_data_poisoning_attack}, (ii) Model Inversion \cite{uav_model_inversion}, and (iii) White-box attacks \cite{uav_white_box_attacks1,uav_white_box_attacks2}. In our research, we specifically address the scenario where an attacker can intercept communication between the drone and its navigation server, posing a potential risk to the UAV's secure operation. Our primary focus is on establishing secure communication channels between the drone and its navigation server, thereby safeguarding it against Targeted Attacks.

Our solution not only mitigates the risk of Targeted Attacks but also protects against Model Inversion attacks. This is achieved by intelligent adaptation of different components of the model architecture to the encrypted domain. The server can be assumed to hold the weights of the model as matrices, and activation functions as polynomial approximations, instead of the true model architecture in sequence. Consequently, even with full knowledge of such weights, an attacker would be unable to configure the architecture, enhancing the security posture of the UAV system. Moreover, the overall execution of the algorithm takes place on encrypted data. Thus one with access to the secret key can only consume the results. However, adversarial image attacks are not protected by this approach. 

\section{FHE basics}
\textbf{Homomorphic encryption (HE) is a cryptographic system that enables computations on encrypted data without the need for decryption, unlike other encryption methods.} In this system, two key components are utilized: public key \(p_k\) and secret key \(s_k\). Encryption and decryption operations are denoted by \(E\) and \(D\), respectively. Consider the plaintext values \(x\) and \(y\), and their corresponding encrypted versions, denoted as \({x'} = E(x, p_k)\) and \({y'} = E(y, p_k)\).

Homomorphic Encryption allows for the computation of various operations directly on encrypted ciphertexts. For instance, the addition of encrypted values (\(x' + y'\)) corresponds to the addition of the original plaintext values (\(x + y\)). Likewise, the multiplication of encrypted values (\(x' * y'\)) is equivalent to the multiplication of original plaintext values (\(x * y\)).

\begin{figure}
\centerline{\includegraphics[width=0.8\linewidth, height=0.4\linewidth]{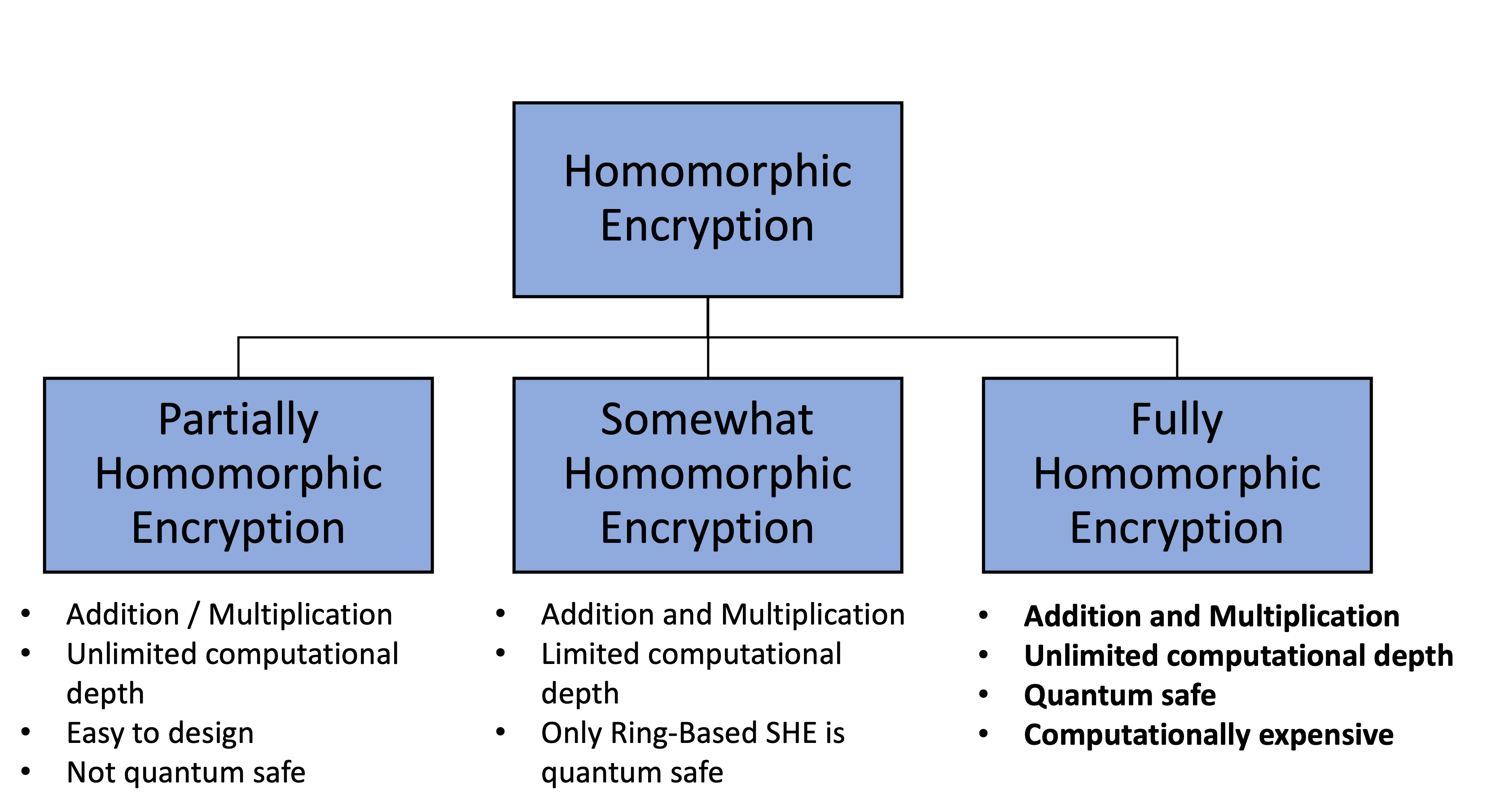}}
    \caption{Types of Homomorphic Encryption (HE) and their features.}
    \label{fig:he_types}
\end{figure}

While there exist various Homomorphic Encryption schemes, \textbf{FHE stands out as the only one capable of supporting computations on ciphertexts of any depth and complexity} as shown in Fig. \ref{fig:he_types}. Various FHE cryptosystems have been proposed - BFV, BGV, and CKKS schemes \cite{gorantala_cacm_fhe}. Notably, BFV and BGV schemes support integers. \textbf{In our research, we have employed the CKKS scheme as it supports floating-point decimals.}

HEAAN, a CKKS FHE scheme, restricts data encryption, allowing only sizes in powers of 2. Hence, we pack our input into arrays of size \(2^n\) before encryption. If the input sizes are not perfect powers of 2, we pad the data with 0s. Although these ciphertexts support Single Instruction Multiple Data (SIMD) operations, they do not provide direct access to individual elements within the ciphertext.

Our research utilizes FHE, specifically the CKKS scheme, to enable secure autonomous UAV navigation using Deep Learning. While FHE allows computations on encrypted data without compromising privacy, certain essential computational operators are yet to be fully implemented in the FHE framework. To address this, we resort to polynomial approximations for these operations. \textbf{In this paper, we have developed FHE-compatible operators tailored for autonomous UAV navigation tasks, leveraging a fully learned deep learning network for inference.}


\section{Related Work}

Numerous surveys have delved into the privacy and security challenges specific to UAVs. Works such as \cite{9795697} and \cite{9488323} highlight the vulnerability landscape in UAV communication networks, emphasizing the delicate trade-off between robust security and the imperative for lightweight, efficient operations. These discussions underscore the crucial role of encryption in fortifying UAV systems against multifaceted threats, as presented by the authors in \cite{8088163}. Our research aims to build upon these foundational insights, contributing to the ongoing discourse on UAV security.

Homomorphic Encryption has been employed in prior work to secure computations in the context of UAV navigation. For instance, in \cite{alzahrani_fhe_uav}, the authors propose an extra key generation encryption technique using the Paillier Cryptosystem to prevent cipher data from being compromised. Further, Cheon et al. \cite{cheon_drone_fhe} explores the development of secure UAVs using a homomorphic public-key encryption method, enabling both secret communication and confidential computation. Another approach focuses on providing a secure and efficient method for third-party UAV controllers to collect and process client data, as demonstrated in \cite{9343124}. The authors propose a Secure Homomorphic Encryption (SHE) framework, which transfers the FHE encryption to UAVs through an encryption protocol.

Despite notable progress in advancing autonomous systems and encryption methodologies for various applications \cite{secure_survey}\cite{LinHAE}\cite{drone_nav_cai}, achieving a comprehensive and practical solution for secure drone systems has proven elusive. While previous works, such as \cite{LinHAE}, offer feasible frameworks for drone controllers, they do not address drone security, leaving them vulnerable to attacks when operating autonomously. Similarly, \cite{drone_nav_cai} presents a secure Reinforcement Learning-based framework for drone navigation, yet its practical implementation remains unfeasible. In contrast to the innovative approach of AutoFHE \cite{auto_fhe} for accelerating inference in encrypted domain of large CNN models (with a focus on ReLU amongst other activations), our work uses a small model with minimal activation functions. 

Among various model compression techniques, including Pruning, Quantization, Decomposition, and Knowledge Distillation \cite{pruning_survey}, our research finds Knowledge Distillation to be particularly effective for FHE. Pruning involves eliminating network components to create sparse models, which, although useful for acceleration and compression, does not significantly reduce computational time for CNNs in FHE. While Quantization typically operates in the BGV scheme, our research focuses on the CKKS scheme \cite{gorantala_cacm_fhe}. Although Decomposition shows promise, it does not match the effectiveness of reducing network depth through Knowledge Distillation.

\begin{figure}[!ht]
\centerline{\includegraphics[width=0.9\linewidth, height=0.35\linewidth]{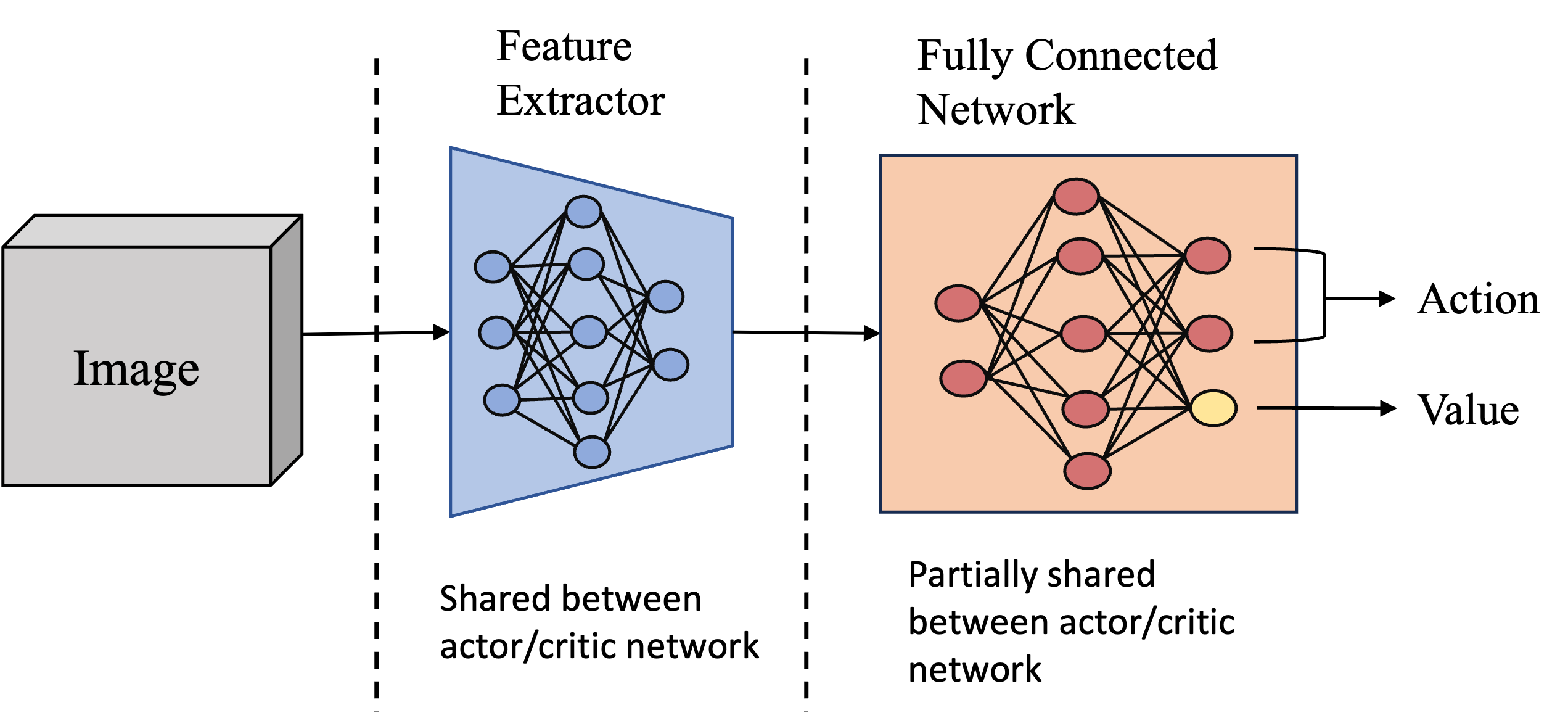}}
\caption{Architecture overview of our framework implementing the Actor-Critic algorithm. }
\label{architecture_overview}
\end{figure}

\section{Proposed Method}

\begin{figure}[!ht]
\centerline{\includegraphics[width=1\linewidth]{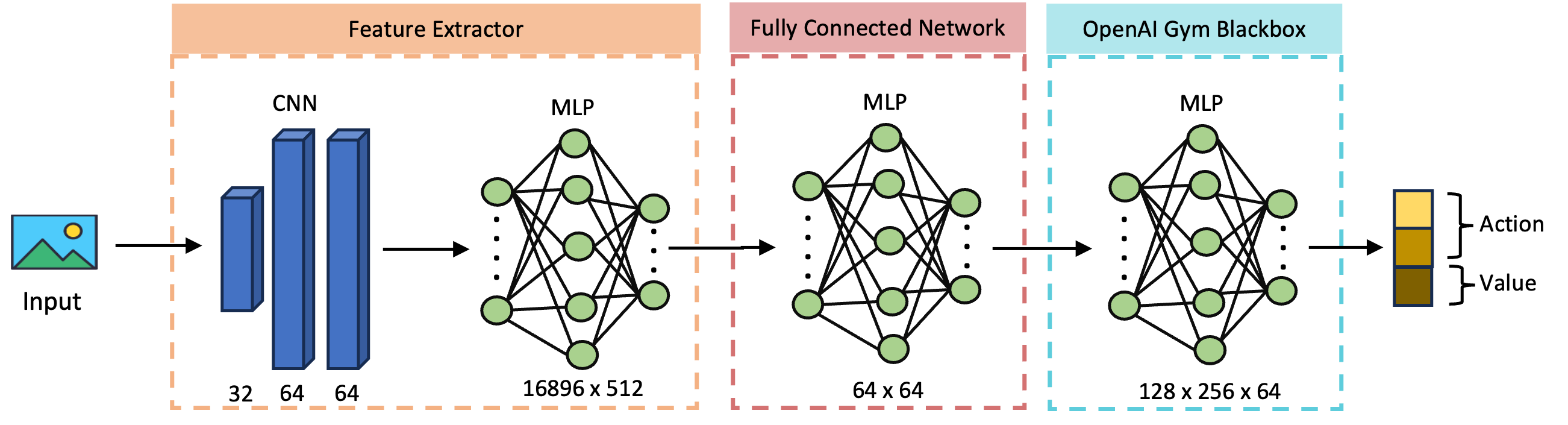}}
\caption{Architecture of the original model (Teacher Network).}
\label{teacher}
\end{figure}

The drone is trained using the Actor-Critic Reinforcement Learning algorithm \cite{schulman2017ppo}. During training, both the Actor and Critic networks are utilized, whereas, during inferencing, only the Actor network is leveraged. The network architecture can be divided into two segments - Feature Extractor and Fully Connected Network as shown in Fig. \ref{architecture_overview}. The Feature Extractor consists of three convolution blocks and one linear block as shown in Fig. \ref{teacher}. Each convolution block consists of a Convolution layer, Batch Normalization layer, and ReLU activation layer. The linear block consists of a Dense Layer, Batch Normalization layer, and ReLU activation layer. The Fully Connected Network segment consists of two shared linear blocks (shared between Actor and Critic) and an output linear block as in Fig. \ref{teacher}. The shared linear blocks are made up of a dense layer and utilize the TanH activation function.

Computation within the Fully Homomorphic Encryption (FHE) domain introduces several significant limitations, including the absence of individual element access in encrypted arrays, restricted computation depth, heightened time complexity, and the absence of inherent support for operators like comparators. Consequently, we choose to train the Actor-Critic model in the unencrypted domain with data generated in a simulated environment, employing Microsoft's AirSim library and Unreal Engine. Subsequently, leverage the model weights for inference within the encrypted domain. To achieve this, we carefully adapt each component of the Actor-Critic network to seamlessly operate within the FHE domain, addressing specific challenges presented by FHE.

In addition to computational constraints, currently, operations in the FHE domain consume significant time. We must have an efficient model with low inference times and high accuracy. We achieve this with the help of Knowledge Distillation in 2 steps. 

Key adaptations within the FHE domain encompass the following components: (i) Model Compression via Knowledge Distillation; (ii) 2-D strided Convolution; (iii) ReLU activation function; (iv) Dense Layer; (v) TanH activation function; and (vi) OpenAI Gym Library. In this section, we provide an in-depth exploration of these adaptations in each layer.

\subsection{Input Adaptations for FHE}

The drone's input comprises of three consecutive images, each captured from the AirSim simulator, with dimensions 50x50. These images are concatenated to form a single input image with dimensions 50x150. In HEAAN, we adopt a strategy where each row of the image is encrypted as a single ciphertext. This approach enables the utilization of SIMD operations, enhancing computational efficiency \cite{9481143}.

Given that HEAAN exclusively supports the encryption of data with sizes as powers of 2, we address this constraint by padding each row of the image with zeros, extending the width to 256. Consequently, the padded input image, now of size 50x256, is encrypted, resulting in a vector of ciphertexts. To facilitate efficient computation, the plaintext weights or filters undergo similar zero-padding, aligning with the dimensions of the padded input image. Importantly, the increase in input size from 50x150 to 50x256 does not impose a significant computational overhead, thanks to the SIMD nature of operations inherent in HEAAN.

\subsection{Knowledge Distillation}

Knowledge distillation, a representative type of model compression and acceleration, effectively learns a small student model from a large teacher model \cite{kd_survey}. In our work, we employ feature-based Knowledge Distillation to compress our original model (Teacher network) to a smaller and FHE-friendly model (Student2 network). We achieve this in 2 steps as shown in Fig. \ref{kd}, achieving Student1 network first and then using Student1 to further compress the model to Student2. It is important to note that, we perform distillation only on the feature extractor network of while training Student1. As shown in Fig. \ref{kd}, we train the student networks on the Cosine Similarity Loss between the extracted features. This significantly reduces the inference time, thereby making the FHE implementation more feasible.

\begin{figure*}[!ht]
\centerline{\includegraphics[width=1\linewidth]{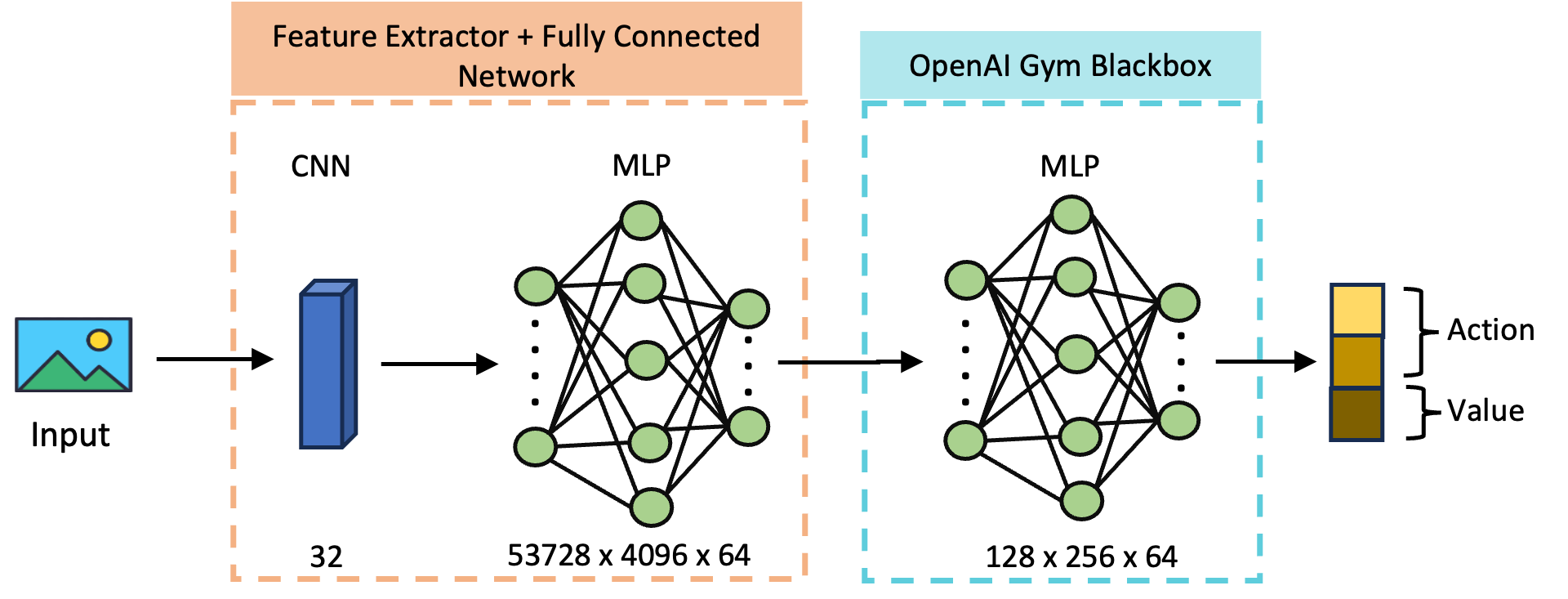}}
\caption{Architecture of the final compressed model (Student2 Netowrk) to comply with FHE's time constraints.}
\label{student2}
\end{figure*}

\begin{figure*}[!h]
\begin{tabular}{ccc}
    
  \begin{tikzpicture}
    \begin{axis}[
      width=4cm,
      height=4.25cm,
      xlabel={Number of filters},
      ylabel={MAE},
      font=\bfseries,
      legend pos=south east,
      legend style ={font=\small},
      xmin=16, xmax=128,
      xmode=log,
      xtick={16, 32, 64, 128},
      xticklabels={16, 32, 64, 128},
      grid=both,
      grid style={line width=.1pt, draw=gray!10},
      major grid style={line width=.2pt,draw=gray!50},
      minor tick num=4,
    ]

    \addplot[mark=*, blue, line width = 0.5pt] table[x=Filter, y=MAE] {
      Filter MAE
      16 0.1724
      32 0.0873
      64 0.0814
      128 0.0791
    };
    \end{axis}
  \end{tikzpicture}
  \label{fig:kd_mae}
&
        
  \begin{tikzpicture}
    \begin{axis}[
      width=4cm,
      height=4.25cm,
      xlabel={Number of filters},
      ylabel={R-squared score},
      font=\bfseries,
      legend pos=south east,
      legend style ={font=\small},
      xmin=16, xmax=128,
      xmode=log,
      xtick={16, 32, 64, 128},
      xticklabels={16, 32, 64, 128},
      grid=both,
      grid style={line width=.1pt, draw=gray!10},
      major grid style={line width=.2pt,draw=gray!50},
      minor tick num=4,
    ]

    \addplot[mark=*, blue, line width = 0.5pt] table[x=Filter, y=R2] {
      Filter R2
      16 0.9073
      32 0.9499
      64 0.9566
      128 0.9578
    };
    \end{axis}
  \end{tikzpicture}
  \label{fig:kd_r2}
  &
  \begin{tikzpicture}
    \begin{axis}[
      width=4cm,
      height=4.25cm,
      xlabel={Number of filters},
      ylabel={Inference Time(s)},
      font=\bfseries,
      legend pos=south east,
      legend style ={font=\small},
      xmin=16, xmax=128,
      xmode=log,
      xtick={16, 32, 64, 128},
      xticklabels={16, 32, 64, 128},
      grid=both,
      grid style={line width=.1pt, draw=gray!10},
      major grid style={line width=.2pt,draw=gray!50},
      minor tick num=4,
    ]

    \addplot[mark=*, blue, line width = 0.5pt] table[x=Filter, y=Time] {
      Filter Time
      16 4205.10
      32 9510.22
      64 14968.49
      128 35027.18
    };
    \end{axis}
  \end{tikzpicture}
  \label{fig:kd_time}\\
    (a) & (b)  & (c)\\

        \end{tabular}
        \caption{(a) Mean Absolute Error (MAE) for various filter counts in the feature-extractor of the Student network (b) R-squared score for various filter counts in the feature-extractor of the Student network (c) Inference time in seconds for various filter counts in the feature-extractor of the Student network.}
        \label{kd_proof}
     
\end{figure*}
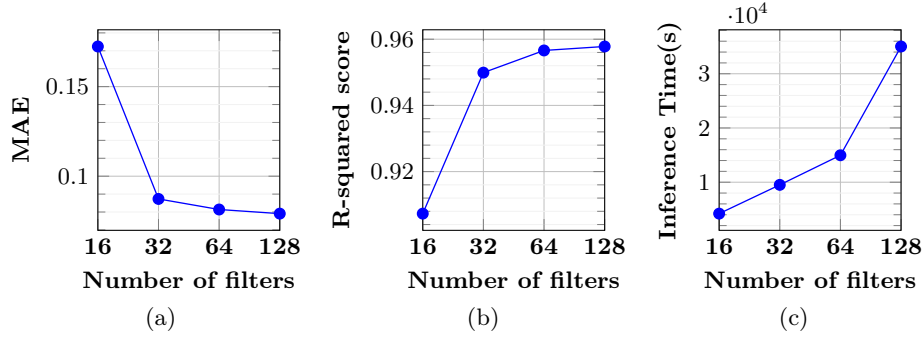

\subsection{Convolutional Layer}

Performing regular convolution in the encrypted domain is extremely computationally inefficient as shown in Table \ref{tab:conv_time} . In our research, we adopt a frequency-domain approach for convolution leveraging the Discrete Fourier transform (DFT). Following are steps performed to achieve 2D convolution in an efficient manner: (i) Perform Homomorphic Fourier Transform (HFT) for each row of 2D Ciphertext using the method in \cite{8701685}; (ii) Take the transpose of 2D Ciphertext using the method proposed in \cite{zekri_transpose}; (iii) Perform row wise HFT of the new transposed Ciphertext; (iv) Transpose back the 2D Ciphertext (v)  Compute the convolution output \(y[n]\) using element-wise multiplication in the frequency domain, as expressed in Equation \ref{eq:convolution}, where \(\mathcal{G}^{-1}\) denotes the inverse Fourier transform, and \(H(u)\) and \(F(u)\) are the DFT of the row of input image and filter, respectively.

\begin{table}[ht]
\centering
\caption{Time complexity analysis of convolution in spatial domain and frequency domain, for an image of size $m\text{x}m$ and filter of size $n\text{x}n$. The time complexities below reflect multiplication complexities.}
\label{tab:conv_time}

\renewcommand{\arraystretch}{1.2}
\begin{tabular}{|c|c|c|}
\hline
\textbf{Convolution domain} & spatial domain & frequency domain \\ \hline
\textbf{Time complexity} & $O(m^2 * n^2)$ & $O(m^2 + 2*n*logn)$\\ \hline
\end{tabular}
\end{table}

\begin{equation}
\label{eq:convolution}
y[n] = \mathcal{G}^{-1}\left\{H(u) \cdot F(u)\right\}
\end{equation}

The DFT of each input value \(h[v]\) is computed using Equation \ref{eq:dft}, where \(H[v]\) represents the DFT coefficient at frequency bin \(v\), and \(N\) is the size of the input.

\begin{equation}
\label{eq:dft}
H[u] = \sum_{v=0}^{N-1} h[v] \cdot e^{-j\frac{2\pi}{N}uv}
\end{equation}

To address the time inefficiency associated with computing the DFT of encrypted data using standard plaintext methods, we employ the Homomorphic Fourier transform. This approach, inspired by Cooley-Tukey matrix factorization \cite{d3ea2d52-5ab2-3128-8b80-efb85267295d}, facilitates an efficient algorithm for computing the 1-D DFT of encrypted data.

\begin{figure*}[!ht]
\centerline{\includegraphics[width=1\linewidth]{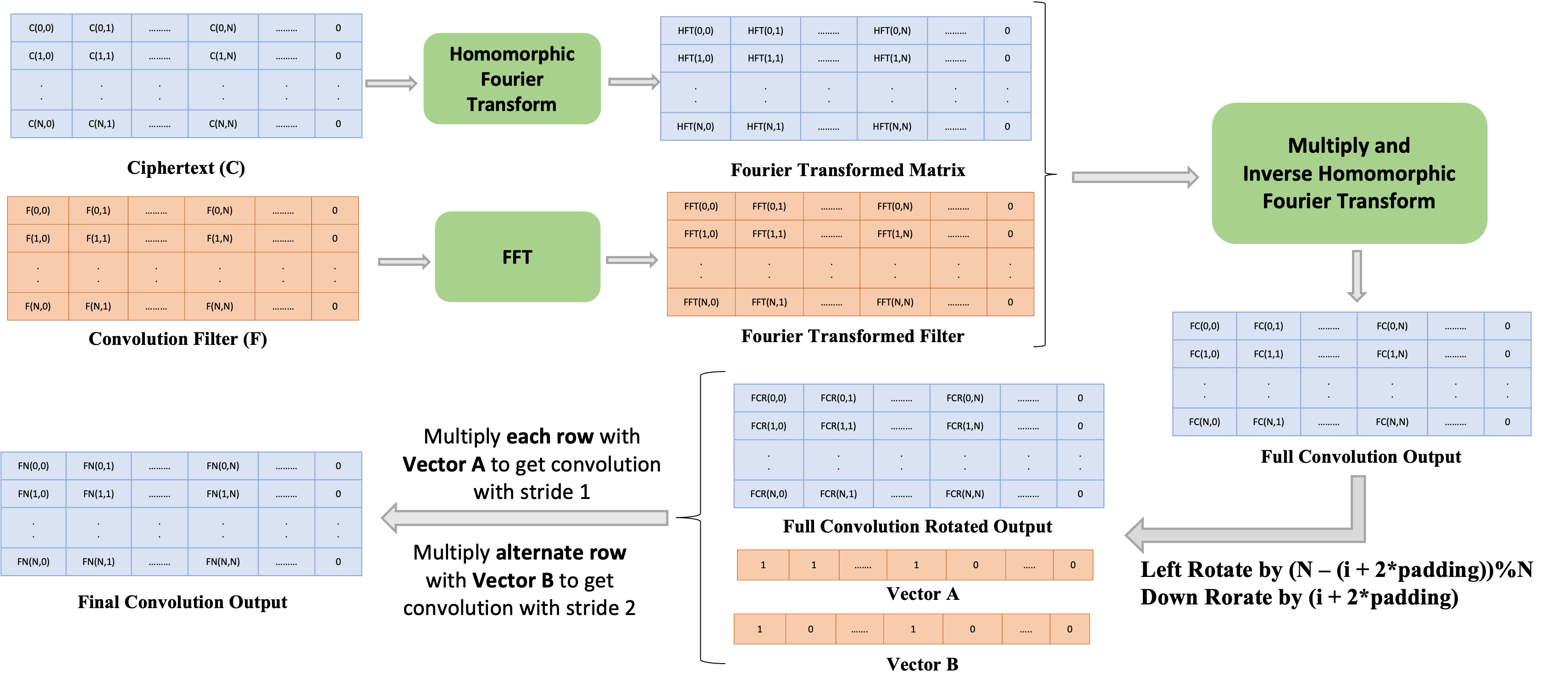}}
\caption{2D Convolution in FHE Domain. Input ciphertext and weights are multiplied in the frequency domain to obtain full convolution. Final convolution
output is obtained by rotating the full convolution as shown above. Different stride-based convolutions can be extracted by multiplying appropriate vectors.}
\label{2d_conv}
\end{figure*}

For transforming the plaintext filter into the frequency domain, we utilize the standard Fast Fourier Transform (FFT). The element-wise multiplication between the input and filter in the frequency domain, followed by the inverse DFT, yields the complete convolution output. To achieve a strided convolution, a rotational manipulation is applied to the resulting ciphertext. We introduce a leftward rotation of the resulting ciphertext by \((N - (2*padding))\%N\) and downward rotation by \(2*padding\), where \(N\) represents the size of the Ciphertext and \(padding\) represents the padded value used to extract DFT convolution output. Additionally, this result is multiplied by an array containing 1s and 0s to obtain appropriate convolution based on the stride value, as illustrated in Fig. \ref{2d_conv}.

\subsection{Activation functions}

Activation functions play a crucial role in neural networks, but their implementation in the context of FHE presents unique challenges \cite{cryptoeprint:2019/417}. FHE libraries lack native support for comparison operations, necessitating the use of approximations like CompG for the sign function \cite{cryptoeprint:2019/1234}. Normalization is essential to align input values within the required range, achieved by scaling the outputs of convolutional layers based on the maximum observed absolute values during training. This scaling factor is determined by the maximum of the absolute values of the inputs' observed range. Following the application of the approximations, positive input values are rescaled to their original range using the inverse of the scaling factor.

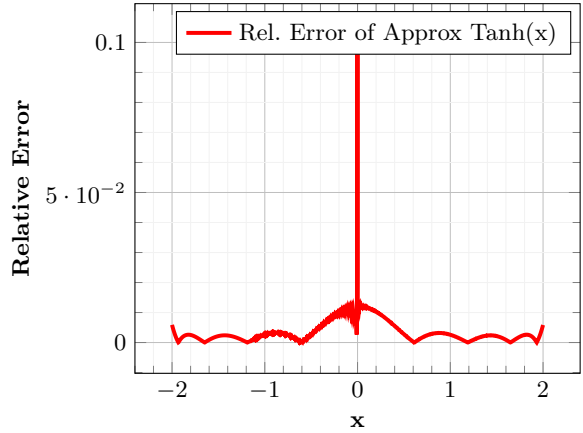
\begin{figure*}[!ht]
\centering
        \begin{tikzpicture}
    \begin{axis}[
      width=7.5cm,
      height=6.5cm,
      xlabel={x},
      ylabel={Relative Error},
      font=\bfseries,
      legend pos=north east,
      legend style={font=\small, at={(0.98,0.98)}, anchor=north east},
      grid=both,
      grid style={line width=.1pt, draw=gray!10},
      major grid style={line width=.2pt,draw=gray!50},
      minor tick num=4,
    ]

    \addplot [
    domain=-2:2, 
    samples=2000,
    color=red,
    line width = 1.5pt
    ]
    {abs((0)*x^8 + (-0.00564857110)*x^7 + (0)*x^6 + (0.0605757714)*x^5 + (0)*x^4 + (-0.279044879)*x^3 + (0)*x^2 + (0.987653369)*x^1 + 0 - ((e^(2*x)-1)/(e^(2*x)+1)))/abs((e^(2*x)-1)/(e^(2*x)+1))};
    \addlegendentry{Rel. Error of Approx Tanh(x)}

    \end{axis}
  \end{tikzpicture}
        \caption{Relative error of f(x) over the interval [-2, 2], where f(x) is the polynomial approximation of Tanh(x). Relative error of f(x) $ = \frac{\left|f(x) - tanh(x) \right|}{\left|tanh(x)\right|}$. }
        \label{rel_err}
     
\end{figure*}

In our research, we adopt a composite approximation technique for comparison in ReLU implementation. This method evaluates the input value \(a\) against zero, encoding the output as 1 for \(a > 0\), 0 for \(a < 0\), and 0.5 for \(a = 0\), and subsequently calculates the final ReLU output by multiplying this result by the input value \(a\). Additionally, we address the challenges of implementing exponential functions in FHE by employing an 8-degree polynomial approximation of TanH restricted to the range [-2, 2]. This approach allows for a closer approximation while mitigating the limitations of FHE in handling exponential functions. The performance of our approximation is evaluated through the relative error of 2000 points within the specified range, providing insights into its effectiveness and accuracy as shown in Fig . ~\ref{rel_err}.

\subsection{Flattening layer}
The flattening operation is usually performed on the convolution outputs. Flattening operation is not possible in FHE without decrypting and re-encrypting the ciphertexts, as it involves changing the length of ciphertexts. To circumvent this issue, we perform element-wise multiplication of the weights and convolution output. Element-wise multiplication is an extremely time-consuming operation as it involves multiplication, addition, and left rotation. We multiply each ciphertext with its corresponding weight vector and add it to a temporary ciphertext initialized to zeros. Then, we perform a summation of the ciphertext elements through repetitive left rotation and addition N-1 times. 

\subsection{Fully-Connected Layer}

A Fully Connected Layer is adapted to FHE as the matrix multiplication of ciphertext inputs and plaintext weight matrices. Each row of weight matrix is multiplied with the ciphertext and the elements of the ciphertext are summed through left rotation.

\subsection{OpenAI Gym Library}
We have adapted the OpenAI Gym Library to FHE through a 3-layer neural network as in Fig. \ref{teacher} and Fig. \ref{student2}. This is due to the limitations of FHE in modeling probability distributions. The neural network learns the probability distribution and maps the final 64-dimension latent vector to the action output. The model is trained in the unencrypted domain and its weights are used for inferencing in FHE.


\section{Results}

Experiments were performed in the encrypted domain on a subset of randomly selected samples from the testing set of the unencrypted domain. We evaluated our results from the FHE-adapted Reinforcement Learning framework against the expected results from the Reinforcement Learning framework in the unencrypted domain. Table \ref{tab:mae} depicts the mean absolute error (MAE) across each block in the Teacher and Student networks within the encrypted domain. Crucially, the regression-based prediction output remained consistent between the FHE version and the plaintext counterpart for the tested samples, indicating coherence in predictive outcomes. We have also achieved an \textbf{R-squared score of 0.9631 for the Teacher network} and \textbf{0.9499 for the Student2 network} with the end-to-end FHE-based Reinforcement Learning framework, in comparison with results in the unencrypted domain. Additionally, Table \ref{tab:layer_time} presents the average processing time across each block in the Teacher and Student networks. We achieve an 18x improvement in inference speed with Knowledge Distillation. These findings substantiate the efficacy of our FHE-adapted network, showcasing the viability of FHE in preserving model accuracy while ensuring data confidentiality.

\begin{figure}[!ht]
\begin{tabular}{cc}
\includegraphics[width=0.49\linewidth]{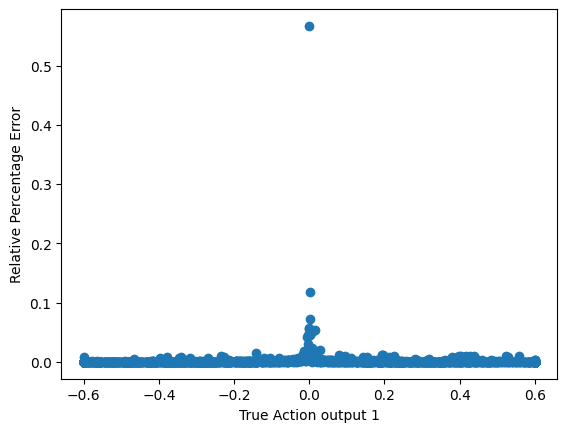} & \includegraphics[width=0.49\linewidth]{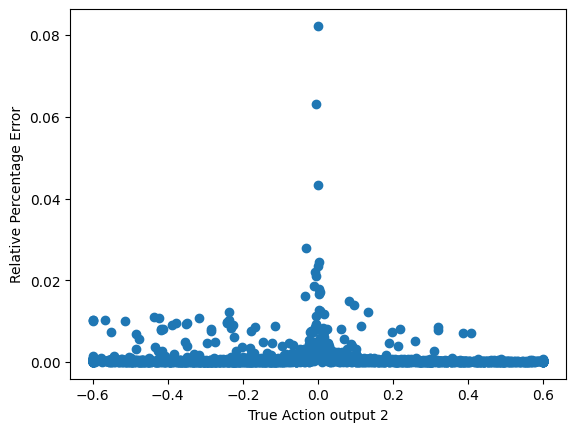}\\
(a) & (b)\\
\end{tabular}
\caption{Relative percentage errors of actions on adaption of OpenAI Gym Library to FHE.}
\label{rel_err_actions}
\end{figure}










\begin{table}[ht]
\centering
\caption{Layerwise average Mean Absolute Error (MAE) between plain-text and FHE model intermediate outputs in Teacher and Student networks.}
\label{tab:mae}

\renewcommand{\arraystretch}{1.5}

\begin{tabular}{|c|r|r|r|}
\hline
\multirow{2}{*}{\textbf{Layer}}       & \multicolumn{3}{|c|}{\textbf{Avergae MAE}}\\ 
\cline{2-4}
& \textbf{Teacher} & \textbf{Student1} & \textbf{Student2} \\ \hline
Convolution  & 0.0779 & 0.0860 & 0.0873 \\ \hline
Linear & 0.0129 & 0.0185 & 0.0203  \\ \hline

OpenAI Gym Library Blackbox & 0.0210 & 0.0206 & 0.0201                  \\ \hline
\end{tabular}
\end{table}







\begin{table}[!h]
\centering
\caption{Time taken by the Teacher and Student networks.}
\label{tab:layer_time}

\renewcommand{\arraystretch}{1.5}

\begin{tabular}{|c|r|r|r|}
\hline
\multirow{2}{*}{\textbf{Layer}}       & \multicolumn{3}{|c|}{\textbf{Inference Time (seconds)}}\\ 
\cline{2-4}
& \textbf{Teacher} & \textbf{Student1} & \textbf{Student2} \\ \hline
Convolution  & 1,006,337.18 & 9,508.44 & 9,510.22      \\ \hline
Linear & 13,662.48 & 43,670.76 &  41,989.52 \\ \hline
OpenAI Gym Library Blackbox  & 4,574.82 & 4,725.92 & 4,668.19 \\ \hline
Total & 1,024,754.48 & 57,905.12 & 56,167.93 \\ \hline
\end{tabular}
\end{table}

\section{Conclusion}

This paper introduces a groundbreaking end-to-end homomorphically encrypted Unmanned Aerial Vehicle (UAV) navigation system, leveraging a fusion of reinforcement learning and deep neural networks. Given Fully Homomorphic Encryption's (FHE) high latency, our results indicate a significant speedup (18x) through Knowledge Distillation. In addition, we seamlessly incorporate convolutional layers, fully connected networks, activation functions, and the OpenAI Gym Library into the FHE domain. The use of the Homomorphic Fourier Transform facilitates efficient convolutions, and an approximate comparator enables the effective mapping of the ReLU activation function. Furthermore, we have devised Tanh approximations, functional mappings from latent feature vectors to action outputs for the Gym Library, and implemented fully connected layers within the FHE domain. In our evaluation of inference, our proposed FHE-based compressed architecture demonstrates lower latency with minimal error across each block in the network, showcasing no discernible accuracy loss when compared to its plaintext counterpart. 


%
%
%
\bibliographystyle{splncs04}
\bibliography{egbib}

\end{document}